\documentclass[pre,twocolumn]{revtex4}
\usepackage{bm}
\usepackage{amsmath}
\usepackage{amsfonts}
\usepackage{amsthm}
\usepackage[pdftex]{graphicx}
\usepackage[utf8]{inputenc}

\newtheorem*{mytheorem}{Theorem}

\begin{document}

\title{Temperature is not an observable in superstatistics}

\author{Sergio Davis}
\homepage{http://www.lpmd.cl/sdavis}
\email{sdavis@cchen.cl}

\affiliation{Comisión Chilena de Energía Nuclear, Casilla 188-D, Santiago, Chile}

\author{Gonzalo Gutiérrez}
\homepage{http://www.gnm.cl/gonzalo}
\email{gonzalo@fisica.ciencias.uchile.cl}

\affiliation{Grupo de Nanomateriales, Departamento de F\'{i}sica, Facultad de Ciencias,
Universidad de Chile, Casilla 653, Santiago, Chile}

\date{\today}

\begin{abstract}
Superstatistics (Physica A \textbf{322}, 267-275, 2003) is a formalism that attempts to explain the
presence of distributions other than the Boltzmann-Gibbs distributions in Nature, typically power-law 
behavior, for systems out of equilibrium such as fluids under turbulence, plasmas and gravitational systems. 
Superstatistics postulates that those systems are found in a superposition of canonical ensembles at different 
temperatures. The usual interpretation is one of local thermal equilibrium (LTE) in the sense of an inhomogeneous 
temperature distribution in different regions of space or instants of time.

Here we show that, in order for superstatistics to be internally consistent, it is impossible to define 
a phase-space function or observable $B(\bm p, \bm q)$ corresponding one-to-one to the local value of 
$\beta=1/k_B T$. Temperature then belongs to a different class of observables than the energy, which has 
as a phase-space function the Hamiltonian $\mathcal{H}(\bm p, \bm q)$.

An important consequence of our proof is that, in Superstatistics, the identification of temperature 
with the kinetic energy is limited to the expectation of $\beta$ and cannot be used to measure the different 
temperatures in LTE or its fluctuations.
\end{abstract}



\maketitle

\section{Introduction}

Superstatistics~\cite{Beck2003, Beck2004} is a relatively new, but already
widely used~\cite{Reynolds2003, Chavanis2006, Porporato2006, Hanel2011} formalism which attempts to 
explain the appearance of non-Boltzmann distributions in Nature for driven or
non-equilibrium systems, and also for small systems~\cite{Dixit2013}. It postulates a weighted superposition 
of canonical ensembles at different temperatures. It has the advantage of not requiring a generalization 
of the entropy functional such as Tsallis' entropy~\cite{Tsallis1988}; it is based solely on the canonical ensemble and the 
correct application of the laws of probability.

Usually superstatistics is understood in terms of fluctuations or inhomogeneities in the physical 
``observable'' corresponding to temperature. It is assumed that there exists a measurable quantity
$T(\bm r, \bm p)$ as a function of positions and \emph{momenta} (a phase-space function) 
which, in principle, could reveal the distribution of temperatures in the system in order 
to characterize it. For a system with Hamiltonian

\begin{equation}
\mathcal{H}(\bm r, \bm p) = K(\bm p) + \Phi(\bm r),
\end{equation}
with $K$ the kinetic energy and $\Phi$ the potential energy, the first candidate that comes to mind is the kinetic temperature,

\begin{equation}
T_K(\bm r, \bm p) = T_K(\bm p) = \frac{2}{3Nk_B}K,
\label{eq_kinetic_estim}
\end{equation}
but there are other possible definitions, based on the so-called dynamical
temperature ~\cite{Rugh1997, Rickayzen2001}. These involve not the momenta but the configurational degrees of freedom. 
A configurational inverse temperature function can be defined as 

\begin{equation}
B_C(\bm r, \bm p) = B_C(\bm r) = \nabla \cdot \left[\frac{{\bm \omega}}{{\bm \omega} \cdot \nabla \Phi}\right]
\end{equation}
where ${\bm \omega}={\bm \omega}(\bm r)$ is a function of position.

In light of the idea of superstatistics, the discussion about the existence of
temperature fluctuations~\cite{Kittel1988, Mandelbrot1989} in thermodynamics has
revived, particularly for the statistical mechanics of small
systems~\cite{Boltachev2010, Falcioni2011, Dixit2013} and because fluctuations
of $\beta$ may be connected to the non-extensivity parameter $q$ in Tsallis statistics~\cite{Beck2003}.

In this work, we show that not only this kinetic temperature $T_K$ fails in its
role as a measure of local or instantaneous temperature in superstatistical
systems but that the problem is deeper: the goal of finding an observable $T(\bm r, \bm p)$ 
with a one-to-one correspondence to the value of temperature $T$ cannot be
achieved. The paper is organized as follows. In Section \ref{sec_statmech}, a
few elements of Statistical Mechanics are reviewed, mainly to fix the notation.
Then in Section \ref{sec_super} these ideas are extended to ensembles with
arbitrary fluctuations of energy. Section \ref{sec_problem} presents the problem
of inferring the underlying ensemble from a set of measurements, and it is in
this context that a hypothetical phase-space function associated to temperature
is postulated. Section \ref{sec_proof} follows with the proof of impossibility
of that function. Finally, Section \ref{sec_conclusions} closes with some
conclusions.

\section{The framework of Statistical Mechanics}
\label{sec_statmech}

Consider a system with degrees of freedom $\bm \Gamma=(\bm r,\bm p)$ and Hamiltonian
$\mathcal{H}(\bm \Gamma)$, whose values we will denote by $E$. This Hamiltonian is bounded 
from below but not from above, i.e., $E_0 < \mathcal{H}(\bm \Gamma) < \infty$. The minimum energy $E_0$ can be set 
to zero without loss of generality. If the system is perfectly isolated so that
its energy is strictly fixed at a value $E$, the probability distribution of the different 
microstates is given by the microcanonical ensemble~\cite{Tuckerman2010},

\begin{equation}
P(\bm \Gamma|E, V, N) = \frac{1}{\Omega(E; V, N)}\delta(\mathcal{H}(\bm \Gamma)-E),
\label{eq_microcanonical}
\end{equation}
where 

\begin{equation}
\Omega(E; V, N) = \int d\bm \Gamma \delta(\mathcal{H}(\bm \Gamma)-E)
\end{equation}
is the density of states. If, on the other hand, the system is placed inside a
heat bath at temperature $T$, the probability distribution of the states is the canonical ensemble,

\begin{equation}
P(\bm \Gamma|\beta) = \frac{\exp\left(-\beta \mathcal{H}(\bm \Gamma)\right)}{Z(\beta)}.
\label{eq_canonical}
\end{equation}
with $\beta=1/k_B T$ and 

\begin{align}
Z(\beta) & = \int d\bm \Gamma \exp\left(-\beta \mathcal{H}(\bm \Gamma)\right) \nonumber \\
& = \int_0^\infty dE \Omega(E)\exp(-\beta E)
\end{align}
the partition function. In order for $Z(\beta)$ to be well-defined, the
temperature $T$ (and therefore $\beta$) cannot be negative. This temperature, in
turn, can be connected with the density of states through the relation

\begin{equation}
\frac{1}{T} = \frac{\partial S(E; V, N)}{\partial E}
\end{equation}
with $S(E; V, N)=k_B \ln \Omega(E; V, N)$ the Boltzmann entropy.

We also know temperature is related to the average kinetic energy of the system through
the equipartition theorem,

\begin{equation}
\Big<\sum_{i=1}^N \frac{p_i^2}{2m_i}\Big>_\beta = \frac{3N}{2}k_B T,
\label{eq_equipartition}
\end{equation}
where $\big<\cdot\big>_\beta$ denote expectation taken over the canonical
distribution with given $\beta$. 

\section{Non-canonical stationary states}
\label{sec_super}

Let us now assume we place the system in a macroscopic stationary state $\mathcal{S}$,
which is neither perfectly isolated nor in equilibrium with a heat bath. In this case, energy will 
fluctuate with a probability distribution $P(E|\mathcal{S})$, and we can always describe the 
new distribution of microstates $P(\bm \Gamma|\mathcal{S})$ as a superposition of microcanonical ensembles 
weighted by $P(E|\mathcal{S})$, that is,

\begin{equation}
P(\bm \Gamma|\mathcal{S}) = \int_0^\infty dE P(E|\mathcal{S})P(\bm \Gamma|E).
\label{eq_decomp_energy}
\end{equation}

Replacing the definition of the microcanonical ensemble (Eq. \ref{eq_microcanonical}), we obtain 

\begin{equation}
P(\bm \Gamma|\mathcal{S}) = \int_0^\infty dE
\left[\frac{P(E|\mathcal{S})}{\Omega(E)}\right]\delta(\mathcal{H}(\bm \Gamma)-E)= \rho(\mathcal{H}(\bm \Gamma)),
\label{eq_rho}
\end{equation}
where we have defined, for simplicity of notation, the function $\rho(E)$ such
that $P(E|\mathcal{S}) = \rho(E)\Omega(E)$. We see that the probability distribution of the microstates 
is a function of the Hamiltonian only, as required by the stationary Liouville
equation,

\begin{equation}
\big\{P({\bf \Gamma}|\mathcal{S}), \mathcal{H}({\bf \Gamma})\big\} = 
\big\{\rho(\mathcal{H}({\bf \Gamma})), \mathcal{H}({\bf \Gamma})\big\} = 0.
\end{equation}

In this case, unlike the microcanonical and canonical ensembles, the ensemble
cannot be described by a single number such as $E$ or $\beta$, instead it can only be described completely 
if we know the shape of the \textbf{function} $\rho$; In this sense we can say
that it is, in fact, a statistical model with an infinite number of parameters.

An alternative to the decomposition in Eq. \ref{eq_decomp_energy} is superstatistics, where $P(\bf \Gamma|\mathcal{S})$ is 
expressed as a superposition of canonical ensembles with different values of $\beta$, that is,

\begin{equation}
P(\bm \Gamma|\mathcal{S}) = \int_0^\infty d\beta P(\beta|\mathcal{S})P(\bm \Gamma|\beta).
\label{eq_model_marg}
\end{equation}
 
Replacing the definition of the canonical ensemble (Eq. \ref{eq_canonical}) and calling $E=\mathcal{H}({\bf \Gamma})$ we have

\begin{equation}
\rho(E) = \int_0^\infty d\beta
\left[\frac{P(\beta|\mathcal{S})}{Z(\beta)}\right]\exp(-\beta E)
\label{eq_beck}
\end{equation}
from which we see that $\rho(E)$ is the Laplace transform of a new function $f(\beta)$ such 
that $P(\beta|\mathcal{S})=f(\beta)Z(\beta)$. This means the function $f(\beta)$ also contains a full 
description of the macrostate $\mathcal{S}$, and for this purpose a determination of $f(\beta)$ is 
equivalent to a determination of $\rho(E)$. We will call these functions the
\emph{ensemble functions}.

It is important to emphasize here the fact that $f(\beta)$ does not correspond
to the probability of observing values of $\beta$, in the same way that
$\rho(E)$ is not the probability of observing the energy $E$. This has somewhat
led to confusion in the literature. The connection between these \emph{ensemble
functions} $f$, $\rho$ and the probability distributions $P(\beta|\mathcal{S})$
and $P(E|\mathcal{S})$ is given by the partition function and density of states, respectively.
A brief summary of this information is given in Table \ref{tbl_properties}.

\section{Can we deduce the stationary ensemble from phase-space measurements?}
\label{sec_problem}

Suppose that we have access to measurements of energy for a particular 
system in a stationary state, and we wish to determine the function $\rho$. We proceed to sample 
$n$ values of energy $E_1, E_2, \ldots, E_n$ and construct an histogram $h$, as

\begin{equation}
h_j = \frac{1}{n}\sum_{i=1}^n \delta(j, k(E_i))
\end{equation}
where $\delta(j,k)$ is Kronecker's delta, $k(E)$ gives the integer position
of the bin corresponding to the value of energy $E$, and $j=1,2,\ldots,m$ with $m$ the total number of bins. 

If $n$ and $m$ are sufficiently large, by the law of large numbers

\begin{equation}
h_j \rightarrow \Big<\delta(E_j-\mathcal{H}({\bf \Gamma}_i))\Big>_\mathcal{S} = P(E_j|\mathcal{S})
\end{equation}
i.e., the histogram will converge to the energy probability distribution
$P(E|\mathcal{S})$, and so, in practice, we can obtain $\rho(E)$ from a large number of energy measurements 
if we know the density of states, as

\begin{equation}
\frac{h_j}{\Omega(E_j)} \approx \rho(E_j).
\label{eq_histogram_energies}
\end{equation}

If we numerically obtain $\rho(E)$ in this way, we could apply the inverse Laplace transform and 
recover the ensemble function $f(\beta)$. But this is redundant because in that case we already would have 
$\rho(E)$, which has all the information to describe the system. We would like a
more direct route to obtain $f(\beta)$, and then the following question arises:

\begin{center}
Is $\beta$ the value of a phase-space function $B(\bm \Gamma)$ in the same way
that $E$ is the value of the Hamiltonian $\mathcal{H}(\bm \Gamma)$?
\end{center}

If such a quantity $B$ exists, and we know the partition function, we can directly obtain 
$f(\beta)$ without the intermediate step of computing $\rho(E)$, just by accumulating enough samples 
$\beta_1=B({\bm \Gamma}_1), \beta_2=B({\bm \Gamma}_2), \ldots, \beta_n=B({\bm \Gamma}_n)$ and the relation

\begin{equation}
\frac{b_j}{Z(\beta_j)} \approx f(\beta_j),
\label{eq_histogram_beta}
\end{equation}
analogous to Eq. \ref{eq_histogram_energies}, where now $b_j$ is the histogram of values $\beta_i$, for 
which the law of large numbers holds as

\begin{equation}
b_j \rightarrow \Big<\delta(B({\bf \Gamma}_i)-\beta_j)\Big>_\mathcal{S}
\end{equation}
and that we can identify with the probability distribution of $\beta$ by

\begin{equation}
P(\beta|\mathcal{S}) = \Big<\delta(B({\bf \Gamma})-\beta)\Big>_\mathcal{S}.
\label{eq_strong_1}
\end{equation}

\begin{table}
\begin{center}
\begin{tabular}{|c|c|c|c|}
\hline
Property & Observable & Ensemble function & Probability density \\
\hline
$E$ & $\mathcal{H}(\bf \Gamma)$ & $\rho(E)$ & $\rho(E)\Omega(E)$ \\
$\beta$ & $B(\bf \Gamma)$ & $f(\beta)$ & $f(\beta)Z(\beta)$ \\
\hline
\end{tabular}
\end{center}
\caption{Features of a superstatistical stationary state $\mathcal{S}$. Note that our 
main result finally shows that there is no suitable definition of the function $B(\bm \Gamma)$.}
\label{tbl_properties}
\end{table}

\section{Impossibility of an intrinsic phase-space function for $\beta$}
\label{sec_proof}

In classical statistical mechanics, we expect that the microscopic observables 
$O$ in our system are defined as phase-space functions $O(\bm \Gamma)$ which are 
independent of the external conditions, being at most functionals of the Hamiltonian 
(which contains all the information about the system and its dynamics). In particular, 
we expect that if we place the system in a stationary \emph{ensemble} $\mathcal{S}$, the
\textbf{definition} of the observable, $O(\bm \Gamma)$, will not change, despite
the fact that its \textbf{value} $\big<O\big>_\mathcal{S}$ most probably will. 
That is, we expect that $O$ is not dependent on the \emph{ensemble} function
$\rho$. This condition can be expressed as 

\begin{equation}
\frac{\delta O(\bm \Gamma)}{\delta \rho(E)} = 0.
\end{equation}

We will call the observables for which this is true, \emph{intrinsic}
observables. They can be defined ``once and for all'' if we know the Hamiltonian
of the system. 

Our main result is that $\beta$ does not fall into this category: there is no intrinsic 
observable $B(\bm \Gamma)$ which gives the superstatistical $\beta$, as shown by the following theorem.

\begin{mytheorem}
In superstatistics, there is no phase-space function $B(\bm \Gamma)$ such that 
$$P(\beta|\mathcal{S}) = \Big<\delta(B(\bf \Gamma)-\beta)\Big>_\mathcal{S},$$
and $$\frac{\delta B}{\delta \rho(E)} = 0.$$
\end{mytheorem}

That is, $B(\bm \Gamma)$ is not an intrinsic observable of the system: even
worse, its definition is dependent on the external conditions that maintain the stationary state, and
thus cannot be used to infer the \emph{ensemble}. In other words, every stationary ensemble 
$\mathcal{S}$ would have its own microscopic definition of temperature.

\begin{proof}
Replacing Eq. \ref{eq_strong_1} into Eq. \ref{eq_model_marg}, we see that

\begin{eqnarray}
\rho(\mathcal{H}(\bm \Gamma)) = \int d\bm \Gamma' \rho(\mathcal{H}(\bm \Gamma'))
\frac{\exp\left(-B(\bm \Gamma')\mathcal{H}(\bm \Gamma)\right)}{Z(B(\bm \Gamma'))}.
\label{eq_condition}
\end{eqnarray}

We can always write the left hand side as 

\begin{equation}
\rho(\mathcal{H}(\bm \Gamma)) = \int d\bm \Gamma' \rho(\mathcal{H}(\bm \Gamma'))\delta(\bm \Gamma'-\bm \Gamma),
\end{equation}
so we have a functional of $\rho$ which is identically zero,

\begin{equation}
\int d\bm \Gamma' \rho(\mathcal{H}(\bm \Gamma'))\left[\delta(\bm \Gamma'-\bm \Gamma)-\frac{\exp\left(-B(\bm \Gamma')\mathcal{H}(\bm \Gamma)\right)}{Z(B(\bm \Gamma'))}\right] = 0.
\end{equation}

Now we will take the functional derivative with respect to $\rho$ on both sides
and assume that $B$ is independent of $\rho$, that is, $\delta B/\delta \rho(E)=0$. It follows that

\begin{equation}
\frac{\exp\left(-B(\bm \Gamma')\mathcal{H}(\bm \Gamma)\right)}{Z(B(\bm \Gamma'))} = \delta(\bm \Gamma'-\bm \Gamma).
\end{equation}

Integrating with respect to $\bm \Gamma'$ we get

\begin{equation}
\int d\bm \Gamma' \exp\left(-B(\bm \Gamma')\mathcal{H}(\bm \Gamma)\right) = Z(B(\bm \Gamma))
\end{equation}
therefore, $B(\bm \Gamma)$ depends on $\bm \Gamma$ only through $\mathcal{H}(\bm \Gamma)$.
Using this, we can write Eq. \ref{eq_condition} as 

\begin{equation}
\rho(E) = \int_0^\infty dE'\Omega(E')\rho(E')\frac{\exp\left(-B(E')E\right)}{Z(B(E'))},
\label{eq_condition_E}
\end{equation}
which again, can be rewritten as

\begin{equation}
\int_0^\infty dE' \rho(E')\left[\delta(E'-E)-\Omega(E')\frac{\exp\left(-B(E')E\right)}{Z(B(E'))}\right] = 0.
\end{equation}

As this must be valid for any $\rho$, we take the functional derivative
$\delta/\delta \rho$ and assume $B$ does not depend on $\rho$. It follows that

\begin{equation}
\Omega(E')\frac{\exp\left(-B(E')E\right)}{Z(B(E'))} = \delta(E'-E),
\end{equation}
for any pair of values $E$ and $E'$, which no function $B(E)$ can fulfill. In order to see why this 
is true, imagine fixing $E'=E_0$ so that $0 < B(E_0) < \infty$. Let us call $\beta_0=B(E_0)$ and $Q=\Omega(E_0)/Z(\beta_0)$. 
Then we have 

\begin{equation}
Q\exp\left(-\beta_0 E\right) = \delta(E_0-E),
\label{eq_cannot}
\end{equation}
for all possible values of $E$. Now, choosing $E=E_0\pm\Delta E$ with $0 <
|\Delta E| < E_0$, we see from Eq. \ref{eq_cannot} that

\begin{equation}
\exp(-\beta_0\Delta E) = \exp(\beta_0\Delta E) = 0,
\end{equation}
which is a contradiction for finite values of $\beta_0$ and $|\Delta E|$. This proves the theorem.

\end{proof}

Despite this result we can provide a useful definition of inverse temperature, 

\begin{equation}
\beta_\mathcal{S} = \Big<\beta\Big>_\mathcal{S}.
\end{equation}
as the \textbf{expectation of the parameter} $\beta$ in the state $\mathcal{S}$. This allows us to
define the temperature as simply $k_B T_\mathcal{S}=1/\beta_\mathcal{S}$. The
inverse temperature $\beta_\mathcal{S}$ can be computed from estimators $\hat{\beta}(\bm r, \bm p)$ 
and this is a value one can use to compare different states or to approximate the 
ensemble to first order by the nearest canonical ensemble. 

In order to show the validity of temperature estimators in an ensemble $P({\bm \Gamma}|\mathcal{S})$
such as the one in Eq. \ref{eq_decomp_energy} (of which superstatistics is a
particular case), we make use of the conjugate variables theorem (CVT)~\cite{Davis2012} for the canonical 
ensemble (a brief proof of which is given in the appendix),

\begin{equation}
\Big<\nabla \cdot {\bm v}\Big>_\beta = \beta\Big<{\bm v}\cdot \nabla \mathcal{H}\Big>_\beta
\label{eq_cvt_canonical}
\end{equation}
and marginalize over $\beta$, using the identity

\begin{equation}
\Big<g(\beta, {\bm \Gamma})\Big>_\mathcal{S} = \int_0^\infty d\beta P(\beta|\mathcal{S})\Big<g(\beta, {\bm \Gamma})\Big>_\beta.
\end{equation}

We see that for the state $\mathcal{S}$ the following CVT holds,

\begin{equation}
\Big<\nabla \cdot {\bm v}\Big>_\mathcal{S} = \Big<\beta {\bm v}\cdot \nabla \mathcal{H}\Big>_\mathcal{S},
\end{equation}
in which $\beta$ is taken as an additional degree of freedom, and the expectation is taken under the joint distribution 
$P({\bm \Gamma}, \beta|\mathcal{S})$. Choosing 

\begin{equation}
{\bm v} = \frac{{\bm \omega}}{{\bm \omega} \cdot \nabla \mathcal{H}}
\end{equation}
as in Ref.~\cite{Davis2012}, we find that

\begin{equation}
\beta_\mathcal{S} = \Big<\hat{\beta}\Big>_\mathcal{S} =
\Big<\nabla \cdot\left[\frac{{\bm \omega}}{{\bm \omega} \cdot \nabla \mathcal{H}}\right]\Big>_\mathcal{S}.
\end{equation}
for any ${\bm \omega}={\bm \omega}({\bm r}, {\bm p})$. It seems suggestive to
associate $\hat{\beta}$ with $B$ but the point of our proof is that precisely,
this choice (or any other) cannot reproduce all the moments of $P(\beta|\mathcal{S})$.

For the particular case of ${\bm \omega}={\bm p}/m$, we obtain a kinetic expression 

\begin{equation}
\beta_\mathcal{S} = \frac{1}{k_B T_\mathcal{S}} = \frac{3N-2}{2}\Big<K^{-1}\Big>_\mathcal{S}
\end{equation}
with $K$ the kinetic energy of the system. Note that, because $\big<K^{-1}\big> > \big<K\big>^{-1}$ 
by Jensen's inequality~\cite{CoverThomas2006},

\begin{equation}
T_\mathcal{S} < \frac{2}{(3N-2) k_B}\Big<K\Big>_\mathcal{S}.
\end{equation}
and so the intuitive generalization of Eq. \ref{eq_kinetic_estim} overestimates the temperature.

\section{Conclusions}
\label{sec_conclusions}

The theorem just proven rules out any intrinsic definition of temperature as a
phase-space function in superstatistics. In statistical terms, we can say that the 
probability distribution $P(\beta|\mathcal{S})$ is not a sampling distribution, and $\beta$ 
has to be interpreted as a parameter.

Our findings do not diminish the power of the superstatistical formalism or
attempt to undermine its foundations. On the contrary, we are led to the
conclusion that the notion of instantaneous or local temperature is at fault and
that it might be separated from the pure idea of superstatistics, keeping $\beta$
as a parameter. There are already efforts to conceptually reformulate
superstatistics from a Bayesian point of view~\cite{Sattin2006}, in which one does 
not need actual variations (temporal or spatial) of a physical quantity. Instead 
there are uncertainties in the well-defined and unique (but unknown) value of $\beta$.

\section{Acknowledgements}
SD gratefully acknowledges funding from FONDECYT 1140514.

\appendix 

\section*{Appendix: Simple proof of the conjugate variables theorem (CVT)}

For an arbitrary distribution of microstates $P(\bf \Gamma)$ let us construct
the expectation of $\nabla \cdot {\boldsymbol \omega({\bf \Gamma})}$, with
$\boldsymbol \omega$ an arbitrary but differentiable vector field,

\begin{equation}
\Big<\nabla \cdot {\boldsymbol \omega}\Big> = \int_V d{\bf \Gamma}P({\bf \Gamma})(\nabla \cdot {\boldsymbol \omega}).
\end{equation}

We consider the divergence theorem applied to a volume $V$ with boundary
$\Sigma$ and ${\bf v} = {\boldsymbol \omega}(\bm \Gamma)P(\bm \Gamma)$,

\begin{equation}
\int_V d{\bf \Gamma}(\nabla \cdot {\bf v}) = \int_\Sigma d{\bf \Sigma} \cdot {\bf v}.
\end{equation}

We obtain 

\begin{align}
\int_V d{\bf \Gamma}\Big[P({\bf \Gamma})\nabla \cdot {\boldsymbol \omega} + {\bf
\omega}\cdot \nabla P({\bf \Gamma})\Big] & = \int_\Sigma d{\bf \Sigma} \cdot
{\boldsymbol \omega}({\bf \Gamma}) P({\bf \Gamma}) \nonumber \\
& = 0,
\end{align}
if the probability $P$ is zero on the boundary $\Sigma$.~\footnote{If this is
not the case, it is always possible to augment the region $V$ to $V'$ and redefine the probability $P$, so 
that $V'$ includes the original integration volume $V$ and assigns zero probability to 
states outside $V$, for instance via a Heaviside step function.}

By replacing $\nabla P$ as $P\nabla \ln P$ we can write both integrals in the lefthand side as expectations 
over $P$, and finally obtain the CVT in its general form,

\begin{equation}
\Big<\nabla \cdot {\boldsymbol \omega}({\bf \Gamma})\Big> + \Big<{\boldsymbol \omega({\bf \Gamma})} \cdot \nabla \ln P({\bf \Gamma})\Big> = 0.
\end{equation}

Replacing $P(\bm \Gamma)$ by $P(\bm \Gamma|\beta)$ in Eq. \ref{eq_canonical} we get the canonical version of CVT, Eq. \ref{eq_cvt_canonical}.

\bibliography{temperature}

\begin{thebibliography}{18}
\expandafter\ifx\csname natexlab\endcsname\relax\def\natexlab#1{#1}\fi
\expandafter\ifx\csname bibnamefont\endcsname\relax
  \def\bibnamefont#1{#1}\fi
\expandafter\ifx\csname bibfnamefont\endcsname\relax
  \def\bibfnamefont#1{#1}\fi
\expandafter\ifx\csname citenamefont\endcsname\relax
  \def\citenamefont#1{#1}\fi
\expandafter\ifx\csname url\endcsname\relax
  \def\url#1{\texttt{#1}}\fi
\expandafter\ifx\csname urlprefix\endcsname\relax\def\urlprefix{URL }\fi
\providecommand{\bibinfo}[2]{#2}
\providecommand{\eprint}[2][]{\url{#2}}

\bibitem[{\citenamefont{Beck and Cohen}(2003)}]{Beck2003}
\bibinfo{author}{\bibfnamefont{C.}~\bibnamefont{Beck}} \bibnamefont{and}
  \bibinfo{author}{\bibfnamefont{E.~G.~D.} \bibnamefont{Cohen}},
  \bibinfo{journal}{Physica A} \textbf{\bibinfo{volume}{322}},
  \bibinfo{pages}{267} (\bibinfo{year}{2003}).

\bibitem[{\citenamefont{Beck}(2004)}]{Beck2004}
\bibinfo{author}{\bibfnamefont{C.}~\bibnamefont{Beck}},
  \bibinfo{journal}{Continuum Mech. Thermodyn.} \textbf{\bibinfo{volume}{16}},
  \bibinfo{pages}{293} (\bibinfo{year}{2004}).

\bibitem[{\citenamefont{Reynolds}(2003)}]{Reynolds2003}
\bibinfo{author}{\bibfnamefont{A.~M.} \bibnamefont{Reynolds}},
  \bibinfo{journal}{Phys. Rev. Lett.} \textbf{\bibinfo{volume}{91}},
  \bibinfo{pages}{084503} (\bibinfo{year}{2003}).

\bibitem[{\citenamefont{Chavanis}(2006)}]{Chavanis2006}
\bibinfo{author}{\bibfnamefont{P.~H.} \bibnamefont{Chavanis}},
  \bibinfo{journal}{Physica A} \textbf{\bibinfo{volume}{359}},
  \bibinfo{pages}{177} (\bibinfo{year}{2006}).

\bibitem[{\citenamefont{Porporato et~al.}(2006)\citenamefont{Porporato, Vico,
  and Fay}}]{Porporato2006}
\bibinfo{author}{\bibfnamefont{A.}~\bibnamefont{Porporato}},
  \bibinfo{author}{\bibfnamefont{G.}~\bibnamefont{Vico}}, \bibnamefont{and}
  \bibinfo{author}{\bibfnamefont{P.~A.} \bibnamefont{Fay}},
  \bibinfo{journal}{Geophys. Res. Lett.} \textbf{\bibinfo{volume}{33}},
  \bibinfo{pages}{L15402} (\bibinfo{year}{2006}).

\bibitem[{\citenamefont{Hanel et~al.}(2011)\citenamefont{Hanel, Thurner, and
  Gell-Mann}}]{Hanel2011}
\bibinfo{author}{\bibfnamefont{R.}~\bibnamefont{Hanel}},
  \bibinfo{author}{\bibfnamefont{S.}~\bibnamefont{Thurner}}, \bibnamefont{and}
  \bibinfo{author}{\bibfnamefont{M.}~\bibnamefont{Gell-Mann}},
  \bibinfo{journal}{Proc. Nac. Acad. Sci.} \textbf{\bibinfo{volume}{108}},
  \bibinfo{pages}{6390} (\bibinfo{year}{2011}).

\bibitem[{\citenamefont{Dixit}(2013)}]{Dixit2013}
\bibinfo{author}{\bibfnamefont{P.~D.} \bibnamefont{Dixit}},
  \bibinfo{journal}{J. Chem. Phys.} \textbf{\bibinfo{volume}{138}},
  \bibinfo{pages}{184111} (\bibinfo{year}{2013}).

\bibitem[{\citenamefont{Tsallis}(1988)}]{Tsallis1988}
\bibinfo{author}{\bibfnamefont{C.}~\bibnamefont{Tsallis}}, \bibinfo{journal}{J.
  Stat. Phys.} \textbf{\bibinfo{volume}{52}}, \bibinfo{pages}{479}
  (\bibinfo{year}{1988}).

\bibitem[{\citenamefont{Rugh}(1997)}]{Rugh1997}
\bibinfo{author}{\bibfnamefont{H.~H.} \bibnamefont{Rugh}},
  \bibinfo{journal}{Phys. Rev. Lett.} \textbf{\bibinfo{volume}{78}},
  \bibinfo{pages}{772} (\bibinfo{year}{1997}).

\bibitem[{\citenamefont{Rickayzen and Powles}(2001)}]{Rickayzen2001}
\bibinfo{author}{\bibfnamefont{G.}~\bibnamefont{Rickayzen}} \bibnamefont{and}
  \bibinfo{author}{\bibfnamefont{J.~G.} \bibnamefont{Powles}},
  \bibinfo{journal}{J. Chem. Phys.} \textbf{\bibinfo{volume}{114}},
  \bibinfo{pages}{4333} (\bibinfo{year}{2001}).

\bibitem[{\citenamefont{Kittel}(1988)}]{Kittel1988}
\bibinfo{author}{\bibfnamefont{C.}~\bibnamefont{Kittel}},
  \bibinfo{journal}{Phys. Today} \textbf{\bibinfo{volume}{41}},
  \bibinfo{pages}{93} (\bibinfo{year}{1988}).

\bibitem[{\citenamefont{Mandelbrot}(1989)}]{Mandelbrot1989}
\bibinfo{author}{\bibfnamefont{B.~B.} \bibnamefont{Mandelbrot}},
  \bibinfo{journal}{Phys. Today} \textbf{\bibinfo{volume}{42}},
  \bibinfo{pages}{71} (\bibinfo{year}{1989}).

\bibitem[{\citenamefont{Boltachev and Schmelzer}(2010)}]{Boltachev2010}
\bibinfo{author}{\bibfnamefont{G.~S.} \bibnamefont{Boltachev}}
  \bibnamefont{and} \bibinfo{author}{\bibfnamefont{J.~W.~P.}
  \bibnamefont{Schmelzer}}, \bibinfo{journal}{J. Chem. Phys.}
  \textbf{\bibinfo{volume}{133}}, \bibinfo{pages}{134509}
  (\bibinfo{year}{2010}).

\bibitem[{\citenamefont{Falcioni et~al.}(2011)\citenamefont{Falcioni,
  Villamaina, Vulpiani, Puglisi, and Sarracino}}]{Falcioni2011}
\bibinfo{author}{\bibfnamefont{M.}~\bibnamefont{Falcioni}},
  \bibinfo{author}{\bibfnamefont{D.}~\bibnamefont{Villamaina}},
  \bibinfo{author}{\bibfnamefont{A.}~\bibnamefont{Vulpiani}},
  \bibinfo{author}{\bibfnamefont{A.}~\bibnamefont{Puglisi}}, \bibnamefont{and}
  \bibinfo{author}{\bibfnamefont{A.}~\bibnamefont{Sarracino}},
  \bibinfo{journal}{Am. J. Phys.} \textbf{\bibinfo{volume}{79}},
  \bibinfo{pages}{777} (\bibinfo{year}{2011}).

\bibitem[{\citenamefont{Tuckerman}(2010)}]{Tuckerman2010}
\bibinfo{author}{\bibfnamefont{M.~E.} \bibnamefont{Tuckerman}},
  \emph{\bibinfo{title}{Statistical {M}echanics: {T}heory and Molecular
  Simulation}} (\bibinfo{publisher}{Oxford University Press},
  \bibinfo{year}{2010}).

\bibitem[{\citenamefont{Davis and Guti\'errez}(2012)}]{Davis2012}
\bibinfo{author}{\bibfnamefont{S.}~\bibnamefont{Davis}} \bibnamefont{and}
  \bibinfo{author}{\bibfnamefont{G.}~\bibnamefont{Guti\'errez}},
  \bibinfo{journal}{Phys. Rev. E} \textbf{\bibinfo{volume}{86}},
  \bibinfo{pages}{051136} (\bibinfo{year}{2012}).

\bibitem[{\citenamefont{Cover and Thomas}(2006)}]{CoverThomas2006}
\bibinfo{author}{\bibfnamefont{T.~M.} \bibnamefont{Cover}} \bibnamefont{and}
  \bibinfo{author}{\bibfnamefont{J.~A.} \bibnamefont{Thomas}},
  \emph{\bibinfo{title}{Elements of Information Theory}}
  (\bibinfo{publisher}{John Wiley and Sons}, \bibinfo{year}{2006}).

\bibitem[{\citenamefont{Sattin}(2006)}]{Sattin2006}
\bibinfo{author}{\bibfnamefont{F.}~\bibnamefont{Sattin}},
  \bibinfo{journal}{Eur. Phys. J. B} \textbf{\bibinfo{volume}{49}},
  \bibinfo{pages}{219} (\bibinfo{year}{2006}).

\end{thebibliography}
\bibliographystyle{apsrev}

\end{document}